\documentclass[preprint,12pt]{iopart}
\usepackage{iopams} 
\usepackage{graphicx} 
\begin{document}

\title[]{Electroosmotic flow rectification in conical nanopores}

\author{Nadanai Laohakunakorn, Ulrich F. Keyser}

\address{Department of Physics, Cavendish Laboratory, JJ Thomson Avenue, \\ Cambridge CB3 0HE, United Kingdom}
\ead{ufk20@cam.ac.uk}
\vspace{10pt}
\begin{indented}
\item[]April 2015 
\end{indented}

\begin{abstract}
Recent experimental work has suggested that electroosmotic flows (EOF) through conical nanopores exhibit rectification in the opposite sense to the well-studied effect of ionic current rectification. A positive bias voltage generates large EOF and small current, while negative voltages generate small EOF and large current. Here we systematically investigate this effect using finite-element simulations. We find that inside the pore, the electric field and salt concentration are inversely correlated, which leads to the inverse relationship between the magnitudes of EOF and current. Rectification occurs when the pore is driven into states characterized by different salt concentrations depending on the sign of the voltage. The mechanism responsible for this behaviour is concentration polarization, which requires the pore to exhibit the properties of permselectivity and asymmetry. 
\end{abstract}

%\pacs{00.00, 00.00, 00.00}
\vspace{2pc}
\noindent{\it Keywords}: nanopore, electroosmotic flow, flow rectification 
%\maketitle

\clearpage

\section{Introduction}

Nanopores are apertures with dimensions of a few tens to a few hundreds of nanometres in an insulating matrix. Immersed in an electrolyte solution, they can conduct a flux of ions and water through their interior. Modulations of this flux by electrically-driven translocation of charged macromolecules form the basis of single-molecule resistive pulse sensing, a major technological application of nanopores today \cite{dekker07,wanunu12}.

Nanopores and nanochannels are characterized by large surface-area-to-volume ratios, and thus the transport of water and ions through these confined geometries is strongly dependent on their surface properties \cite{schoch08,sparreboom09}. In particular, nanopores made from silica or glass typically develop a negative surface charge at neutral pH \cite{behrens01}. When in contact with a salt solution, the surface is screened by an electric double layer a few nanometres thick containing mobile positive counterions. Application of a tangential electric field results in an electric force on this screening layer. Viscous transfer of momentum away from the surface leads to bulk fluid motion known as electroosmotic flow (EOF) \cite{happel83}. 

In addition to fluid flow, the screening layer affects ionic transport significantly as the dimensions of the system are reduced. As the conductivity due to the double layer becomes comparable to the conductivity of the pore's lumen, the pore exhibits permselectivity towards cations \cite{stein04,steinbock12}. 

In this regime of surface-governed transport, interesting effects have been observed. As with permselective ion-exchange membranes, driving an ionic current through a permselective pore results in concentration polarization (CP) across the pore \cite{kim07}. In order for flux balance to be achieved, the total salt concentration reduces upstream of the permselective region, and increases downstream \cite{zangle10}. As the driving voltage is increased, the upstream depletion becomes more pronounced until the supply of charge carriers is exhausted, leading to a limiting current behaviour. At higher voltages still, electrokinetic instabilities result in vortices which destroy the depletion region, and conduction is recovered through the pore \cite{kim07,rubinstein00}. 

If an asymmetry is introduced along the axial direction of the pore, for instance by making the pore conical in structure, the conductive properties of the pore now change with respect to voltage reversal. This is because one polarity of the applied voltage results in a depletion region inside the pore, while the other results in enrichment. The pore therefore exhibits ionic current rectification, and behaves like an ionic diode \cite{siwy04,siwy06,woermann03,constantin07}. 

The ionic transport properties of rectifying nanopores have been well-studied. However, in addition to ionic current rectification, there has been recent experimental evidence suggesting that electroosmotic flow (EOF) through conical nanopores also exhibits interesting and partly unexpected behaviour \cite{jin10,laohakunakorn13_1,laohakunakorn15}. In particular, EOF has been observed to exhibit flow rectification which behaves in the opposite sense to current rectification, i.e. large currents are correlated with small EOF, and \emph{vice versa}. The relative scarcity of studies on EOF rectification in single nanopore systems is most likely due to difficulties in measuring the effect, as compared with ionic current measurements.   

Rectification of EOF has potentially significant consequences as it could guide the development of nanofluidic flow rectifiers. These would play the same role in microfluidic circuits as electronic diodes do in electrical circuits. Due to the linearity, and hence time-reversal symmetry, of the Stokes equations governing low Reynolds number fluid flow, engineering low Reynolds number flow rectifiers has traditionally proven to be a challenging problem \cite{groisman04,sousa10}. 

In this paper, we present the results of finite-element simulations which capture the rectification behaviour of electroosmotic flows in a nanopore. We show that the fluxes of water and ionic current through the pore depend on the local electric field and salt concentration. By comparing our results with simple analytic models, we can predict the relationship between field and concentration inside the pore. Finally, we demonstrate how concentration polarization leads to variations in these quantities, resulting in rectification of EOF and ionic current which are intricately linked.

\section{Model and Methods}

\begin{figure} 
\centering 
\includegraphics[width=0.5\textwidth]{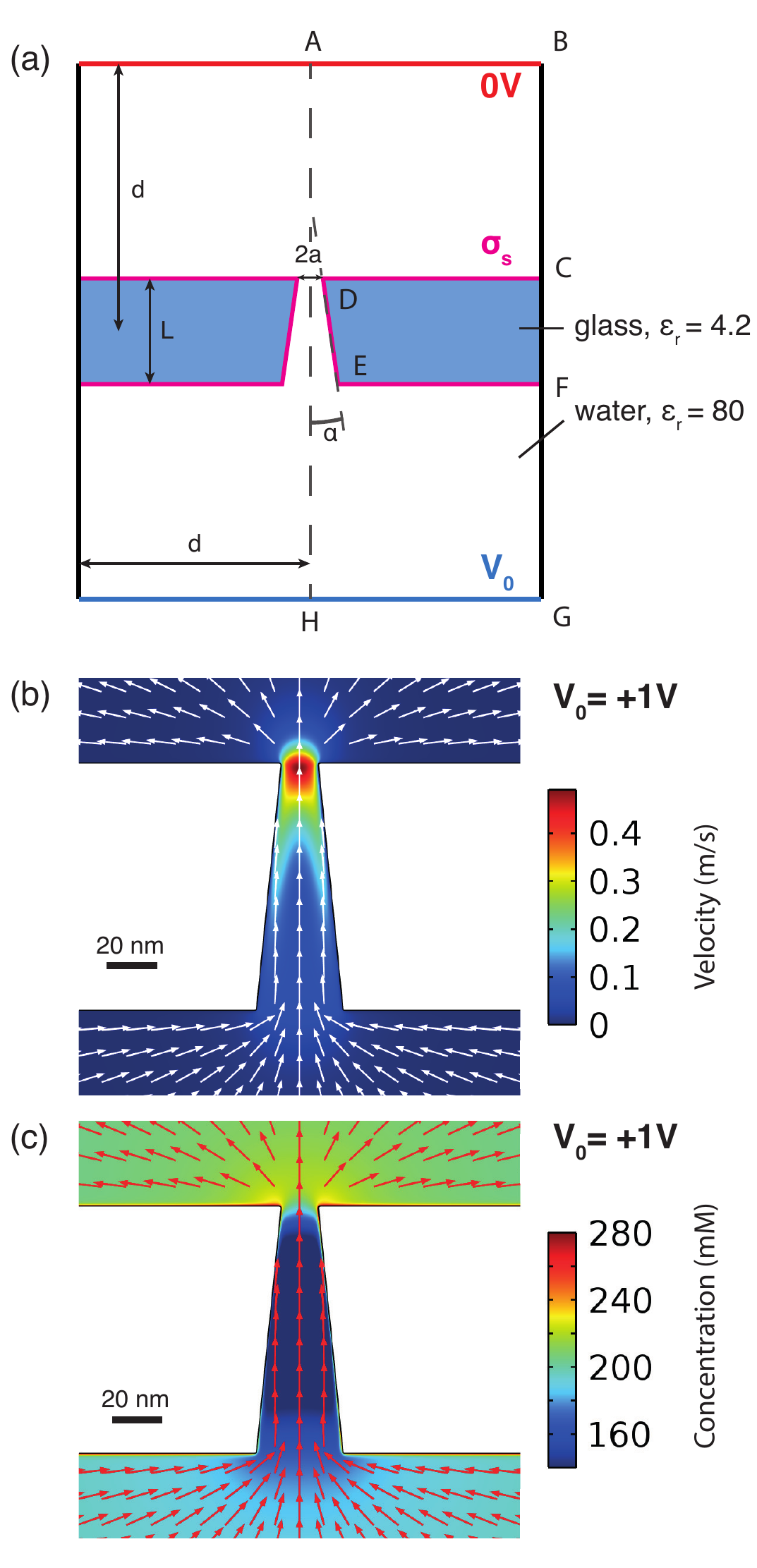}
\caption{Finite-element simulations of transport through a nanopore. (a) The simulation geometry is a 2D axisymmetric design about the axis AH. The pore is modelled as a cone of angle $\alpha$ and tip diameter $2a$ embedded in a membrane of thickness $L$ whose surface CDEF contains a charge density $\sigma_s$. The box size $d$ is chosen such that it mimics infinitely large reservoirs. A voltage is applied between two electrodes AB and HG. (b) Typical flow profiles through a pore with $\alpha=0.1$ rad for an applied voltage of $V_0=+1$ V. Arrows indicate flow direction, and colours indicate flow magnitude. (c) Typical ionic flux through the pore for the same voltage. Arrows indicate the direction of net ionic current, while colours represent the total salt concentration. The bulk salt concentration is $2c_0=200$ mM.} 
\label{fig:geometry} 
\end{figure}

Finite-element simulations have proven to be a powerful technique in the analysis of nanopore physics \cite{laohakunakorn15,mao14,mao13,ai10}. For pores with the smallest dimensions larger than a few nm, the physics of ion and fluid transport is well-captured by classical continuum equations. Compared to atomstic and molecular dynamics simulations, continuum simulations can deliver accurate, quantitative results at a fraction of the computational cost, as long as the pore dimensions are sufficiently large that coarse-graining of molecular details is possible. We use the commercial software package COMSOL Multiphysics 4.4 to carry out finite-element calculations.

Our simulation geometry is shown in Figure 1a. It is a 2D-axisymmetric geometry which can be revolved about the axis AH to produce full 3D solutions. The pore has a radius $a=7.5$ nm and is embedded in a membrane of length $L=100$ nm. The taper of the pore is parameterized by an angle $\alpha$. The simulation is enclosed within a box of size $d\sim10L$, which is large enough to mimic infinity (Supplementary Information). The membrane is modelled as a solid with permittivity $\epsilon_r\sim4.2$, similar to that of quartz glass; the solution is water with permittivity $\epsilon_r\sim80$. The surface of the membrane contains a constant surface charge density of $\sigma_s$. The constant $\sigma_s$ approximation is reasonable at moderate salt concentrations ($\sim100$ mM) \cite{vanderheyden05}. 

We use a free triangular mesh with a nine-level boundary layer at the surface CDEF. The thickness of the first boundary layer is set to $\sim\lambda_D/10$, where $\lambda_D$ is the Debye screening length, to ensure the electric double layer is adequately resolved. The Debye length is the characteristic thickness of the electric double layer \cite{schoch08}. The mesh is heavily refined inside the pore as well as near the corners C, D, E, and F (Figure 1a).

Transport through nanopores is well-described within the Poisson-Nernst-Planck-Stokes (PNPS) formalism \cite{schoch08}. The electric potential $\phi(\boldsymbol{r})$ is related to the charge density $\rho_e(\boldsymbol{r})$ by the Poisson equation:
\begin{equation}
\bnabla^2\phi(\boldsymbol{r})=-\frac{\rho_e(\boldsymbol{r})}{\epsilon_0\epsilon_r}, \label{Poissontheory}
\end{equation}
where $\epsilon_0\sim8.85\times10^{-12}$ Fm$^{-1}$ is the permittivity of free space. The ionic flux $\boldsymbol{J}_i$ of species $i$ is calculated using the Nernst-Planck equation:
\begin{equation}
\boldsymbol{J}_{i}=-D_i\bnabla c_i(\boldsymbol{r})-\frac{D_iz_ie}{k_BT} c_i(\boldsymbol{r})\bnabla\phi(\boldsymbol{r})+c_i(\boldsymbol{r})\boldsymbol{u}(\boldsymbol{r}) \label{NernstPlancktheory},
\end{equation}
where $D_i$, $c_i(\boldsymbol{r})$, and $z_i$ are the diffusion constant, concentration, and valency of species $i$ respectively, $e\sim1.6\times10^{-19}$ C is the elementary charge, $k_B\sim1.38\times10^{-23}$ JK$^{-1}$ is the Boltzmann constant, $T=300$ K the temperature, and $\boldsymbol{u}(\boldsymbol{r})$ the fluid velocity field. For KCl, $z_i=\pm1$ and $D_{K^+}\sim D_{Cl^-}\sim2\times10^{-9}$ m$^2$s$^{-1}$. Finally, the fluid velocity is obtained using the Stokes equation:
\begin{equation}
\mu\bnabla^2\boldsymbol{u}=\bnabla p +\rho_e(\boldsymbol{r})\bnabla\phi(\boldsymbol{r}), \label{Stokestheory}
\end{equation}
where $\mu$ is the dynamic viscosity and $p$ the pressure inside the fluid.

Our simulation proceeds in two steps. First, a 1D solution is obtained along the boundary BCFG, with the boundary conditions $\phi=0$ at B, $\phi=V_0$ at G, surface charge $\sigma_s$ at C and F, and $c_i=c_0$ at B and G. This results in solutions for the electric potential $\phi_{1D}$ and concentrations $c_{i, 1D}$ for both species along the boundary. The pressure $p_{1D}$ can be obtained from the Stokes equation by substituting in the solution for $\phi_{1D}$. 

In the second step, the PNPS equations are solved fully coupled in the 2D geometry using the boundary conditions $\phi=0$ on AB, $\phi=V_0$ on HG, surface charge $\sigma_s$ on CDEF, and on AB and HG, no fluid stress, and $c_i=c_0$. The solutions $\phi_{1D}$, $c_{i, 1D}$, and $p_{1D}$ are used as Dirichlet boundary conditions on BCFG. Our method is very similar to that used in a previous study \cite{mao14} to model induced-charge electroosmosis through nanopores. 

We run the simulations for $c_0=100$ mM, and varying $V_0$, $\sigma_s$, and $\alpha$. The solutions are a set of fields $\phi(\boldsymbol{r})$, $c_i(\boldsymbol{r})$, $\boldsymbol{u}(\boldsymbol{r})$, and $p(\boldsymbol{r})$. Typical flow fields and concentration profiles are shown in Figure 1b and c. In order to obtain the ionic current $I$ and flow rate $Q$, we carry out the following integrals:
\begin{equation}
I=N_A e\int_S (\boldsymbol{J}_{K^+}-\boldsymbol{J}_{Cl^-})\cdot\boldsymbol{dS} \label{currentfluxint}
\end{equation}    
\begin{equation}
Q=\int_S \boldsymbol{u}\cdot\boldsymbol{dS} \label{flowfluxint},
\end{equation}    
where $S$ is the surface area spanning the cross-section of the pore. For our calcuations, we chose the location of $S$ to be halfway through the pore. Full details of the simulation procedure are given in the Supplementary Information. 

\section{Results}

\subsection{Rectification of ionic current and EOF}
\begin{figure} 
\centering 
\includegraphics[width=0.5\textwidth]{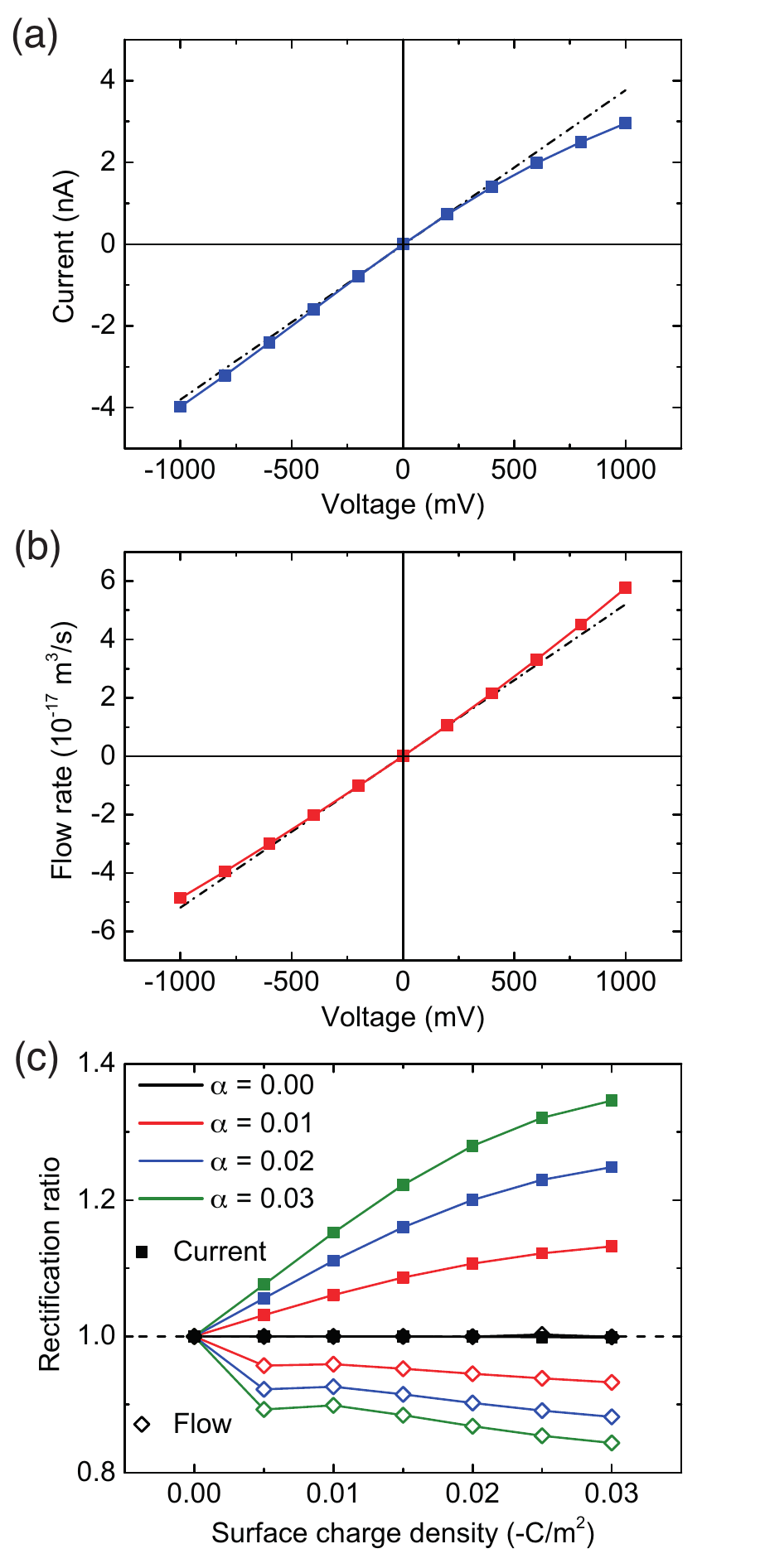}
\caption{Rectification of ionic current and electroosmotic flow. (a) Squares indicate ionic current, and the dashed line is a fit about $V_0=0$ V. The current grows sublinearly at positive voltages, and superlinearly at negative voltages. Here, $\alpha=0.03$ rad, and $c_0=100$ mM. (b) A plot of the flow rate for the same pore, showing the rectification behaviour in the opposite sense. (c) The rectification ratio, defined as $r_{I}=|I(-1V)/I(+1V)|$ for current (filled squares) and $r_{Q}=|Q(-1V)/Q(+1V)|$ for the flow (open diamonds), is plotted as as function of surface charge density for different taper angles (measured in radians). Rectification increases with surface charge, and also with taper angle. The conditions required for rectification therefore are a non-zero surface charge as well as a non-zero taper angle.} 
\label{fig:rectification} 
\end{figure}

Figure 2a and b show typical variations of the ionic current and flow rate as a function of applied voltage, for a negatively-charged pore. Linear fits about $V_0=0$ are shown by the dashed lines. For positive biases, the magnitude of the current is less than would be expected from the linear relationship, while the magnitude of EOF is greater. The reverse is true for negative biases. This behaviour we define as rectification, which can be quantified by defining rectification ratios as follows:
\begin{equation}
r_{I}=\left|\frac{I(-1V)}{I(+1V)}\right|
\end{equation}
\begin{equation}
r_{Q}=\left|\frac{Q(-1V)}{Q(+1V)}\right|
\end{equation}
where the definition has been chosen to be consistent with previous literature. For a negatively-charged pore, therefore, $r_I>1$ , while $0<r_Q<1$.

Figure 2c shows the current and flow rectification ratios as a function of surface charge, for different values of taper angle. There are three main observations from these results. First, current and flow rectification behave in the opposite sense, as observed experimentally \cite{laohakunakorn13_1,laohakunakorn15}. Second, the degree of rectification depends on the surface charge, with more highly-charged pores exhibiting greater rectification. A non-zero surface charge is therefore a necessary condition for rectification. The final observation is that rectification additionally requires an asymmetry such as that introduced by a non-zero taper angle. The greater the pore's asymmetry, the greater the rectification. The aim of this paper is to develop an understanding of these observations. 

\clearpage
\subsection{Comparison with analytic infinite cylinder model}

\begin{figure} [h]
\centering 
\includegraphics[width=1\textwidth]{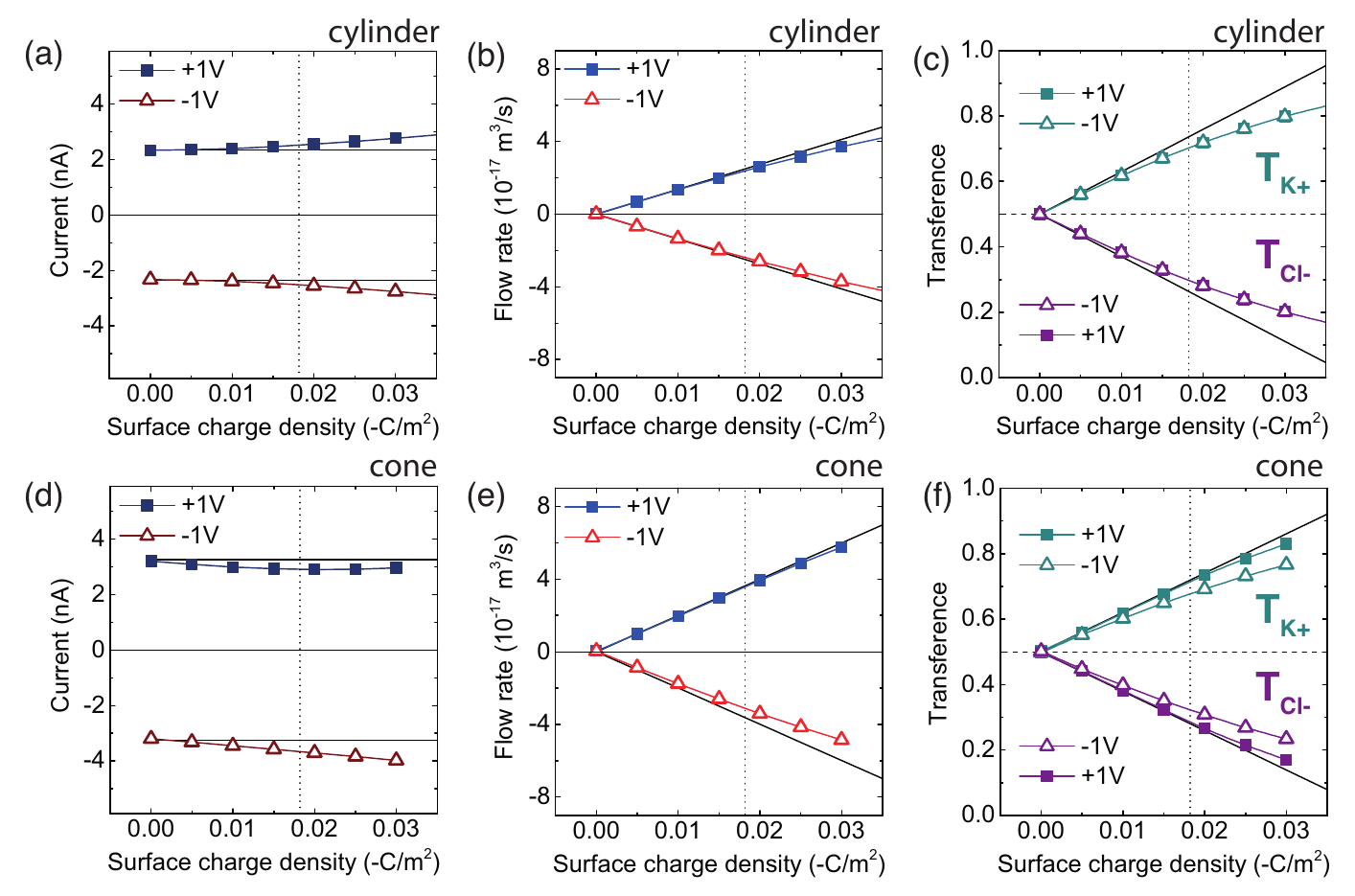}
\caption{Simulated flux values through the pore. (a--c) The simulated current, flow rate, and transference through a cylindrical pore (symbols), and the analytic calculations of the same quantities through an infinite cylinder of equal radius (solid black lines). Quantities at positive voltages are represented by filled squares, and those at negative voltages by open triangles. The transference for both cations (cyan) and anions (purple) are shown; the sum of the transference is always 1. Vertical dotted lines indicate the limit of validity of the Debye-H\"{u}ckel approximation used to derive the analytic results. (d--f) The same as above, but for a cone of $\alpha=0.03$ rad. In this case, the magnitudes of the quantites are not preserved under voltage reversal. The analytic calculations are now carried out for an infinite cylinder of the same effective radius (equal to the average radius of the cone). } 
\label{fig:flux} 
\end{figure}

The factors which influence the current and flow can be understood by considering a simple model consisting of an infinite cylindrical channel under a constant applied electric field. For such a geometry, standard analytic solutions for the ion distribution and EOF velocity are available under the Debye-H\"{u}ckel approximation $|e\phi/(k_B T)|\leq 1$ \cite{muthukumar11}. It is therefore possible to analytically determine the integrals in equations \ref{currentfluxint} and \ref{flowfluxint}. For small surface charges, we obtain the following linearized expressions for ionic current and flow rate (derivation in Supplementary Information):
\begin{equation}
I = \pi a^2D\epsilon\kappa^2E_z \label{Ilin}
\end{equation}
\begin{equation}
Q = -  \frac{\pi a^2}{\mu \kappa}\frac{I_2(\kappa a)}{I_1 (\kappa a)}E_z\sigma_s, \label{Qlin}
\end{equation}
where $a$ is the cylinder radius, $E_z$ the axial electric field, $\kappa=\sqrt{2N_A e^2 c_0/(k_B T \epsilon_0\epsilon_r)}$ is the inverse Debye length, $\epsilon=\epsilon_0\epsilon_r$, and $I_n$ is the $n^{th}$-order modified Bessel function of the first kind. The inverse Debye length $\kappa\propto\sqrt{c_0}$ is a parameter which characterizes the salt concentration, and in the rest of this paper the ionic strength will be referred to in terms of $\kappa$.

In addition, an important quantity to consider is the transference number $T_i$, defined as the proportion of total current carried by an ionic species $i$: $T_i=I_i/I$. The linearized cation transference number is given by (Supplementary Information):
\begin{equation}
T_{K^+}=\frac{1}{2}+\frac{1}{2}\left[\frac{k_BT}{De\mu\kappa}\left(\frac{2}{\kappa a}-\frac{I_0(\kappa a)}{I_1(\kappa a)}\right)-\frac{2e}{\epsilon k_B T\kappa^2 a}\right]\sigma_s. \label{Tlin}
\end{equation}  
These equations show that at small surface charges, $I$ is independent of the surface charge, while $Q$ and $T_{K^+}$ are linear in $\sigma_s$. Figure 3 compares simulated values of $I$, $Q$, and $T_{K^+}$ with these analytic expressions. We observe that the linearized infinite cylinder model is an excellent approximation for the finite cylinder at small surface charges, as expected (Figure 3a--c). As surface charge is increased, the current does not vary to first order, while the flow rate increases linearly. The cationic transference also increases, showing clearly the increasing permselectivity of the pore.

Once a taper angle is introduced, the infinite cylinder model still captures the qualitative trends in $I$, $Q$, and $T_{K^+}$, with one important exception: the simulated results show that the magnitude of these quantities is no longer the same upon voltage reversal (Figure 3d--f). At negative voltages, the magnitude of $I$ is increased compared to at positive voltages, while $Q$ and $T_{K^+}$ are smaller. 

Equations \ref{Ilin}, \ref{Qlin}, and \ref{Tlin} suggest that the only quantities that can change upon voltage reversal are electric field and salt concentration, since all other quantities remain constant. Because the equations do not determine $E_z$ and $\kappa$, requiring instead that they are specified by other means, they cannot explicitly capture the rectification behaviour. However, the main conclusion that can be drawn is the fact that the change in magnitudes of $I$, $Q$, and $T_{K^+}$ upon voltage reversal are due to a change in electric field and salt concentration.

\subsection{Relationship between electric field and salt concentration}
\begin{figure} [h]
\centering 
\includegraphics[width=1\textwidth]{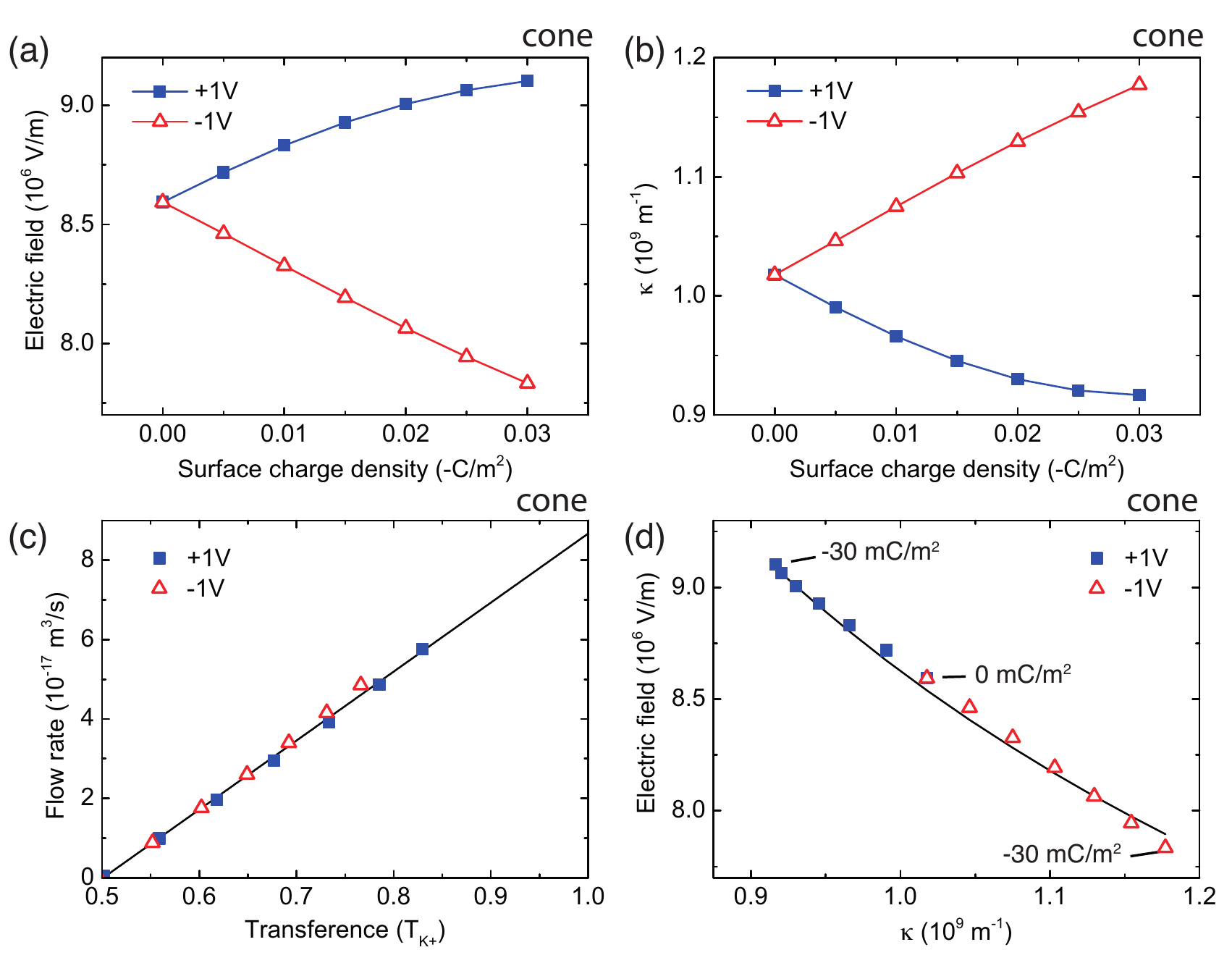}
\caption{Relationship between electric field and salt concentration. (a) The average electric field magnitude inside a conical pore with $\alpha=0.03$ rad as a function of surface charge density. For positive biases, as charge is increased, the field magnitude increases, while at negative voltages, the field decreases in magnitude. (b) The average salt concentration (measured in terms of $\kappa$, the inverse Debye length) exhibits the opposite behaviour. As charge is increased, concentration decreases at positive voltages, but increases at negative voltages. (c) The magnitude of the flow rate is a linear function of transference, and the slopes of the relationship at positive and negative voltages are roughly equal. This sets a constraint for the relationship between the electric field and salt concentration. (d) The electric field magnitude and concentration inside the pore are anti-correlated. Negative voltages tend to drive the pore into a low field, high concentration state, while positive voltages result in a high field, low concentration state. The solid line is the analytic relationship expressed in equation \ref{const}.} 
\label{fig:fieldsalt} 
\end{figure}

In order to investigate the relationship between electric field and salt concentration, we begin by calculating the average value of these quantities over the entire volume of the pore. Figure 4a and b show the average electric field magnitude and salt concentration (measured in terms of the inverse Debye length $\kappa$) inside the conical pore, as a function of surface charge, for voltages of $V_0=\pm1$ V. Here, the difference in behaviour due to the sign of the applied voltage becomes apparent. In Figure 4a, we see that for a given surface charge, positive voltages result in high electric fields, while negative voltages result in low fields. The opposite effect is true for the salt concentration: positive voltages result in a lower salt concentration compared to negative voltages. The variations in electric field and salt concentration are thus anti-correlated. 

We can investigate the dependence of electric field on concentration in more detail by considering the relationship between the magnitude of flow rate and transference, as plotted in Figure 4c. This shows that the flow is a linear function of transference. The intuition behind this observation is that, for symmetric electrolytes with equal mobilities, cationic transference is essentially a measure of the prevalence of cations over anions, and hence the degree of momentum imbalance in the fluid. Without an imbalance, i.e. for $T_{K^+}=T_{Cl^-}=0.5$, there can be no net force and hence no EOF. By eliminating $\sigma_s$ between equations \ref{Qlin} and \ref{Tlin}, one obtains 
\begin{equation}
Q=\frac{\pi a^2  e }{\frac{ e^2\mu}{\epsilon k_B T}+\left(\frac{k_B T}{D}\right)\left(\frac{\kappa a}{2} \frac{I_0(\kappa a)}{I_1(\kappa a)}-1\right)}\frac{ I_2(\kappa a)}{I_1 (\kappa a)}(\kappa a) E_z\left[T_{K^+}-\frac{1}{2}\right]. \label{QT}
\end{equation}
The observed linear relationship in Figure 4c, and the nearly equal slopes for both positive and negative applied voltages, indicate that the prefactor in equation \ref{QT} is approximately constant. It is important to point out that \emph{a priori}, there is no reason to expect that the slopes should be equal upon voltage reversal. The consequence of this observation is that the variation in electric field and salt concentration is constrained to obey
\begin{equation}
\frac{\pi a^2  e }{\frac{ e^2\mu}{\epsilon k_B T}+\left(\frac{k_B T}{D}\right)\left(\frac{\kappa a}{2} \frac{I_0(\kappa a)}{I_1(\kappa a)}-1\right)}\frac{ I_2(\kappa a)}{I_1 (\kappa a)}(\kappa a) E_z\approx\mathcal{C}, \label{const}
\end{equation}
where $\mathcal{C}= Q/T_{K^+}$. Figure 4d shows the electric field against salt concentration, and the analytic relationship given by equation \ref{const}. This plot clearly shows the constraint on the variation of electric field and salt concentration inside the pore: low fields are correlated with high concentrations, and vice versa. 

An intuitive understanding of the anti-correlated behaviour of electric field and salt concentration can be obtained by considering Ohm's law, which states that $J=\sigma E$, where $\sigma$ is the conductivity of the solution. For a constant current, as the conductivity reduces, the electric field increases. This would result in a constraint $E\propto\kappa^{-2}$, i.e. the same as that expressed by equation \ref{Ilin}. However, it is important to note that under rectification conditions, the current is no longer constant. Since equation \ref{const} takes into account varying current as well as EOF, it is thus a more sophisticated constraint than Ohm's law.

\subsection{The relationship between ionic current and EOF}
\begin{figure} [h]
\centering 
\includegraphics[width=1\textwidth]{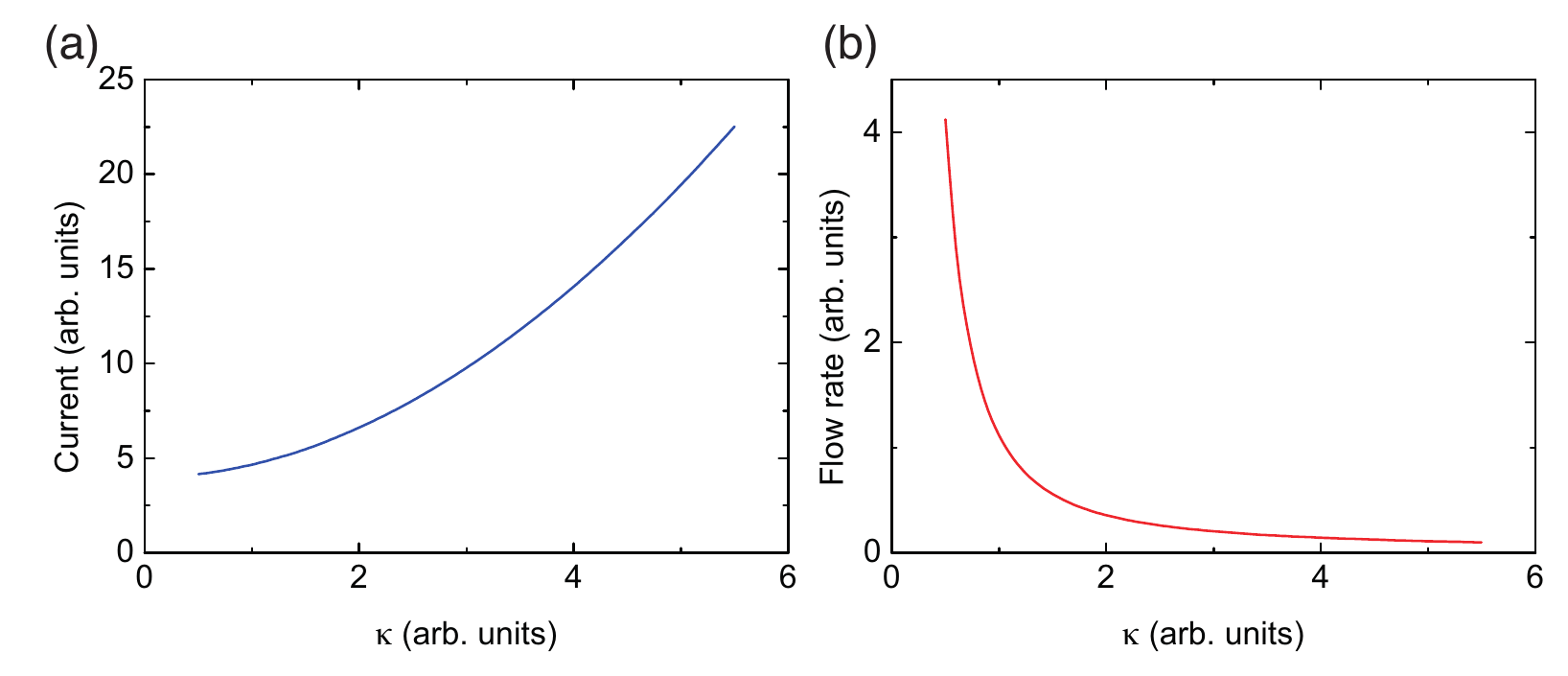}
\caption{The variation of current and EOF with salt concentration. (a) The current is an increasing function of salt concentration, as described by equation \ref{Ik}. Values plotted here are in arbitrary units. (b) The flow rate is a decreasing function of salt concentration, as described by equation \ref{Qk}.} 
\label{fig:currentflow} 
\end{figure}

Under situations where the constraint given by equation \ref{const} is valid, the ionic current and flow rate can be written down purely in terms of the salt concentration.
\begin{equation}
I=\mathcal{A}\; \mathcal{C}\; \frac{\kappa I_1(\kappa a)}{I_2 (\kappa a)}\left[1+\Xi\left(\frac{\kappa a}{2}\frac{I_0(\kappa a)}{I_1(\kappa a)}-1\right)\right] \label{Ik}
\end{equation}
\begin{equation}
Q=\mathcal{B}\; \mathcal{C}\;\frac{1}{\kappa^2}\left[1+\Xi\left(\frac{\kappa a}{2}\frac{I_0(\kappa a)}{I_1(\kappa a)}-1\right)\right], \label{Qk}
\end{equation}
where the constants $\mathcal{A}=De\mu/(k_B T a)$, $\mathcal{B}=-\sigma_s e/(\epsilon k_B T a)$, and $\Xi=\epsilon(k_B T)^2/(De^2\mu)$. Only one $(E,\kappa)$ pair is needed to set the constant $\mathcal{C}$, and hence in these expressions the functional dependance of the electric field is removed completely. 

Figure 5 shows the form of these two equations. Whereas the ionic current is an increasing function of salt concentration, the flow rate decreases as concentration increases. A change in salt concentration thus leads to variation of current and flow in opposite directions. 

Whilst we have described the variations of electric field and concentration in detail, we have made no mention yet of the cause of these variations: in particular, what is required is a mechanism which results in different concentrations inside the pore as a function of voltage.  

\subsection{Concentration polarization and symmetry breaking}
\begin{figure} [h]
\centering 
\includegraphics[width=1\textwidth]{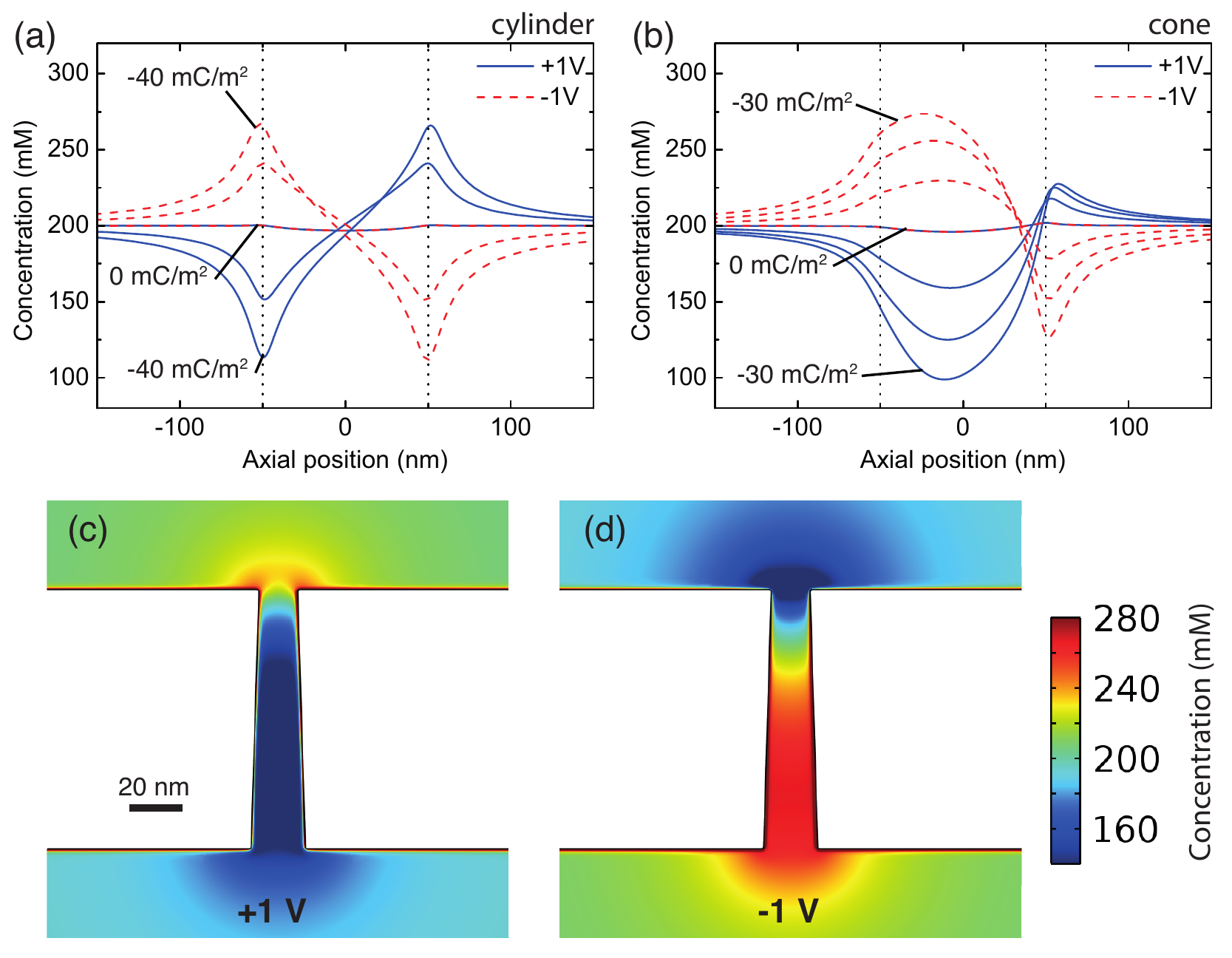}
\caption{Concentration polarization in the nanopore. (a) The variation of salt concentration along the axis of the pore, for voltages of $\pm1$ V applied from left to right; vertical lines indicate the extent of the pore. An ionic current flowing through a permselective pore results in a depletion region upstream, and an enrichment region downstream of the pore. For a cylinder, the polarization is symmetric about the centre of the pore. Different lines correspond to different values of surface charge density, which is increased in equal intervals from 0 to -40 mC/m$^2$. (b) When the pore contains a taper angle of $0.03$ rad, the CP is asymmetric with respect to the centre of the pore. At positive voltages, the upstream depletion region is extended inside the pore compared to the enrichment region outside. (c--d) The spatial variation in concentration near the same conical pore (with $\sigma_s=-0.03$ C/m$^2$), showing clearly the depletion at positive voltages and enrichment at negative voltages.} 
\label{fig:CP} 
\end{figure}

As discussed earlier, an ionic current flowing through a permselective pore results in a depletion region upstream, and an enrichment region downstream of the pore, an effect called concentration polarization (CP). CP arises due to the requirements of flux balance. Within the permselective region, the majority of the cationic flux is due to electromigration; however, upstream of this region, the same flux is distributed between a smaller electromigration component, and a diffusive component towards the pore. The diffusive flux results from a concentration gradient which decreases approaching the pore, which characterizes the upstream depletion region.  

Conversely, downstream of the pore the flux is distributed between an electromigration component and a diffusive component away from the pore, which results from a concentration gradient decreasing away from the pore. This characterizes the downstream enrichment region.

Electroneutrality requires that the anionic concentrations are depleted and enriched in the same way as the cations, and thus the variation applies to the overall salt concentration, not just that of a single species.

In a cylindrical pore, this polarization is symmetric with respect to the centre of the pore (Figure 6a). However, by introducing a taper angle, the polarization can be made asymmetric with respect to the direction of current flow: depletion and enrichment regions inside the pore become much larger than the corresponding regions outside (Figure 6b). Hence, the entire pore can be driven into a depleted, low-concentration state under positive voltages, and a high-concentration state under negative voltages (Figures 6c and d). This asymmetric behaviour is the origin of the rectification effects.    

The final picture we have for current and flow rectification can therefore be summarized as follows: in a permselective, conical pore, a positive voltage leads to CP which reduces the salt concentration inside the pore. This reduction in salt concentration puts the pore in a low-conductivity state, and hence a low ionic current flows. At the same time, the low concentration is correlated with high electric fields and a high cationic transference number, as cations form a larger proportion of the total ionic concentration. The high electric fields and prevalence of cations over anions result in large electroosmotic flows. The converse, when a negative voltage is applied, results in a high conductivity, low transference, and low flow state for the pore. 

Because CP is absent in a non-permselective pore, the presence of surface charge is a necessary condition for rectification. In order to obtain asymmetric CP, an asymmetric pore is required, which is the second necessary condition. Our model thus correctly predicts the conditions required for rectification as presented in Figure \ref{fig:rectification}c.  

\section{Discussion}
We have presented a unified description of ionic current and electroosmotic flow rectification. The study has been guided by analytic relations derived from an infinite cylinder model, and is therefore valid for small surface charges, small taper angles, and relatively large aspect ratio pores (which are long compared to their width). These requirements have motivated the conditions we have used: surface charges of 0 to -0.03 C/m$^2$ compare well with literature values of charges on glass and silica surfaces \cite{behrens01}, and are small enough that the Debye-H\"{u}ckel approximation is valid for a reasonable range of values. We have presented results for taper angles between 0--0.03 radians, which corresponds to a maximum total opening angle of $\sim3^{\circ}$. Experimentally, conical glass nanopores can have opening angles up to $\sim10^{\circ}$ \cite{laohakunakorn15}. Finally, our model describes conical pores with a length of around 100 nm, and can be best applied to systems such as conical glass nanopores \cite{laohakunakorn13_1}. Since typical solid-state nanopores are made in membranes with thicknesses on the order of a few tens of nm \cite{li01}, we expect effects of the membrane outside these pores to play a more important role \cite{laohakunakorn15,lee12}, and hence our description of the pore as a long cylinder will require modification when applied to these systems. 

As the parameters vary outside these ranges, we expect a departure from linearity. This is already exhibited in the increasing current, and sublinear growth in flow and transference (e.g. Figure 3a--c) at high surface charges. This behaviour is due to the salt concentration inside the pore being effectively dominated by the ions inside the double layer. As the pore becomes more highly charged, electroneutrality requries that the total charge in the double layer balances the total charge on the walls, and so the salt concentration inside the pore slowly increases beyond even the bulk values. This leads to higher ionic current and a slower growth in the flow, as expected. 

Although we have presented results for fixed pore radius and salt concentration, an increase in either radius or salt concentration will diminish rectification effects, as observed experimentally (e.g. \cite{laohakunakorn15}). This can be attributed to the decreased permselectivity of the pore in both cases.

In the Supplementary Information, we show results for charge densities up to -0.08 C/m$^2$ and taper angles of $\alpha=0.5$ rad, as well as the effects of varying the pore length and radius. In this case, although the linear approximations break down, the results preserve their qualitative properties, and we believe that as long as the system remains free from instabilities, such as those associated with overlimiting current behaviour \cite{rubinstein00}, our conclusions will remain valid. 

As mentioned earlier, it is intuitive that flow rate and transference should be correlated, as the cationic transference measures the charge imbalance which leads to a net force in the first place. However, the observation that the relationship between the two quantities is linear (Figure 4c) is a stronger statement, which allows us to write down the constraint given by equation \ref{const}. The origin and generality of this constraint will be investigated in future work.

Finally, although in our studies the asymmetry which is responsible for rectification is geometric, we expect that any other property which breaks the $(-z\rightarrow z)$ symmetry of the system will lead to similar EOF rectification effects. An example would be a spatially-varying surface charge, which has already been demonstrated to lead to current rectification in cylindrical pores \cite{karnik07}, as well as the closely-related effect of \emph{osmotic} flow rectification \cite{picallo13}.    

\section{Conclusions}
We have carried out an investigation into the rectification of electroosmotic flows in a conical nanopore. Like current rectification, this effect requires the pore to possess both permselectivity and geometrical asymmetry. The flux of current and water through the pore are determined by the electric field and salt concentration inside the pore, two quantities which we have found to be anti-correlated. Permselectivity and asymmetry result in the pore exhibiting concentration polarization: when a positive voltage is applied at the wide end of the pore, the pore is driven into a low concentration state associated with low current and high electric fields, transference, and EOF. A negative voltage drives the pore into a high concentration state, which results in high currents and low electric fields, transference, and EOF. The rectification of current and flow is therefore intricately linked.

\ack
The authors wish to acknowledge Dr. Sandip Ghosal (Northwestern University) for useful discussions in the early stage of this work, as well as Dr. Stefan Kesselheim (University of Stuttgart) and Mao Mao (Northwestern University) for their guidance on the COMSOL simulations. UFK and NL acknowledge funding from an ERC starting grant, Passmembrane 261101. 

\section*{References}

\bibliography{bibliography_full}
\bibliographystyle{unsrt}

\end{document}